\documentclass{iaus}
\usepackage{graphicx}

\def\Msolar{\hbox{${\rm M}_\odot$}}

\title[Have loose globular clusters collapsed yet?] 
{Have loose globular clusters collapsed yet?}

\author[De Marchi, Paresce \& Pulone]   
{Guido De Marchi$^1$, Francesco Paresce$^2$ \and Luigi Pulone$^3$}

\affiliation{$^1$ESA, Space Science Department, 2200 AG Noordwijk, Netherlands 
\break email: gdemarchi@rssd.esa.int \break 
$^2$INAF, Istituto di Astrofisica Spaziale e Fisica Cosmica, 40129 
Bologna, Italy
\break email: paresce@iasfbo.inaf.it \break
$^3$INAF, Observatory of Rome, 00040 Monte Porzio Catone, Italy 
\break email: pulone@mporzio.astro.it}

\pubyear{2008}
\volume{246}  
\pagerange{119--126}
\date{17.9.2007}
\jname{Dynamical Evolution of Dense Stellar Systems}
\editors{E. Vesperini, M. Giersz, A. Sills}
\begin{document}

\maketitle

\begin{abstract}
We report on the discovery of a surprising observed correlation between
the slope of the low-mass stellar global mass function (GMF) of
globular clusters (GCs) and their central concentration parameter
$c=\log(r_t/r_c)$, i.e. the logarithmic ratio of tidal and core radii.
This result is based on the analysis of a sample of twenty Galactic GCs,
with solid GMF measurements from deep HST or VLT data, representative of
the entire population of Milky Way GCs. While all high-concentration 
clusters in the sample have a steep GMF, low-concentration clusters tend 
to have a flatter GMF implying that they have lost many stars via 
evaporation or tidal stripping. No GCs are found with a flat GMF and high 
central concentration. This finding appears counter-intuitive, since the 
same two-body relaxation mechanism that causes stars to evaporate and the 
cluster to eventually dissolve should also lead to higher central density 
and possibly core-collapse. Therefore, severely depleted GCs should be in 
a post core-collapse state, contrary to what is suggested by their low 
concentration. Several hypotheses can be put forth to explain the observed
trend, none of which however seems completely satisfactory. It is likely
that GCs with a flat GMF have a much denser and smaller core than 
suggested by their surface brightness profile and may well be undergoing 
collapse at present. It is, therefore, likely that the number of post 
core-collapse clusters in the Galaxy is much larger than thought so far.


\end{abstract}


The dynamical evolution of globular clusters (GCs) is governed by the
two-body relaxation process, whereby stars exchange energy via repeated
distant encounters (see Spitzer 1987 and Elson, Hut \& Ingaki 1987 for a 
review). Two-body encounters lead to the expansion of the outer regions 
of the cluster, while driving the stellar density in the central regions 
to increase dramatically towards an infinite value during the so-called 
core-collapse. The most visible effect of core-collapse is the appearence
of a central cusp in the surface brightness profile of the cluster 
(Djorgovski \& King 1986) and, correspondingly, an increase in the
central concentration parameter $c=\log(r_t/r_c)$, i.e. the logarithmic 
ratio of tidal and core radii. Therefore, $c$ has traditionally been
seen as a gauge of the dynamical state of a cluster, with values
of $c$ in excess of $\sim 2$ indicating a post-core-collapse phase 
(Djorgovski \& Meylan 1993; Trager, Djorgovski \& King 1995).

Besides driving a cluster towards core-collapse, equipartition of
energy through two-body relaxation also alters over time the mass 
distribution. More massive stars tend to transfer kinetic energy
to lighter objects and sink towards the cluster centre, while less 
massive stars migrate outwards. The resulting mass segregation implies
that the local stellar mass function (MF) within a cluster changes 
with time and place. Even if the stellar initial mass function (IMF) 
was the same everywhere when the cluster formed (a condition that does
not seem to be true in some very young rich clusters where more massive
stars are already more centrally concentrated; see e.g. Sirianni et al.
2002), after a few relaxation times there will be proportionately more 
low-mass stars in the cluster periphery and proportionately less in the 
core. The first tentative evidence of mass segregation in 47 Tuc (Da 
Costa 1982) has been fully confirmed by early HST observations in this
and other clusters (see e.g. Paresce, De Marchi \& Jedrzejewski 1995; 
De Marchi \& Paresce 1995) and is now acknowledged in all observed GCs. 

Finally, another important effect of the two-body relaxation process
is that it drives the velocity distribution towards a Maxwellian and, 
therefore, an increasing number of stars in the tail of the velocity 
distribution will acquire enough energy to exceed the escape velocity 
and leave the cluster. This phenomenon, called evaporation, happens even 
in an isolated cluster, but its extent is greatly enhanced by the 
presence of the tidal field of the Galaxy, in a way that depends on the 
cluster's orbit. Evaporation, coupled with tidal truncation, is the 
leading cause of mass loss for most GCs after the first few billion years
of their formation (see e.g. Gnedin \& Ostriker 1997) and causes a 
preferential loss of low-mass stars, since these have typically higher 
velocities. The result is a selective depletion at the low-mass end of the 
cluster's stellar global MF (GMF). This effect, integrated over the orbit 
and time, implies a progressive departure of the GMF from the stellar IMF
(Vesperini \& Heggie 1997), namely a flattening at low masses that is now 
well established observationally (De Marchi et al. 1999; Andreuzzi et al.
2001; Koch et al. 2004; De Marchi, Pulone \& Paresce 2006; De Marchi \& 
Pulone 2007).

It would, therefore, appear natural to expect that, as clusters evolve
dynamically, the increase in their central concentration should correspond
to the flattening of their GMF, as both effects result from the same two-body 
relaxation process. To test this hypothesis, we have built a sample of 20 GCs 
for which reliable estimates exist of both $c$ and the shape of the GMF (see 
De Marchi, Paresce \& Pulone 2007 for details). The former comes from 
accurate surface photometry (Harris 1996), while the latter has been
determined by us using high-quality HST and VLT photometry as briefly
explained here below.    

In order to obtain the GMF of a cluster, one would need to measure the MF
of its entire stellar population, since MF measurements limited to a specific 
location need to be compensated for the effects of mass segregation. We have 
shown, however, that if the MF is measured at various locations inside the 
cluster (e.g. near the core, at the half-light radius and in the periphery), 
the GMF can effectively be derived by constraining a model MF to reproduce
simultaneously three observables: the radial variation of the MF, the
surface brightness profile and the velocity dispersion profile. Details on 
how this is done, using multi-mass Michie--King models, can be found in De 
Marchi et al (2006) and references therein. Our analysis of clusters with 
an almost complete radial coverage of the MF (De Marchi et al. 2006; De 
Marchi \& Pulone 2007) proves the long cherished belief that the local MF 
near the half-light radius is for all practical purposes indistinguishable 
from the GMF (Richer et al. 1991; De Marchi \& Paresce 1995; De Marchi, 
Paresce \& Pulone 2000).

In this work we have limited our analysis of the GMF to the mass range 
$0.3 - 0.8$\,\Msolar, in which a power-law distribution of the type $dN/dm
\propto m^\alpha$ appears to adequately reproduce the observations. This
choice of the mass range is dictated by the fact that below
$\sim 0.3$\,\Msolar\, the GMF of GCs departs from a simple power-law (see
Paresce \& De Marchi 2000 and De Marchi, Paresce \& Portegies Zwart 2005)
and, more importantly, because the number of clusters with reliable 
photometry at those masses is still limited.

\begin{figure}
\centering
\includegraphics[height=11cm,width=12cm]{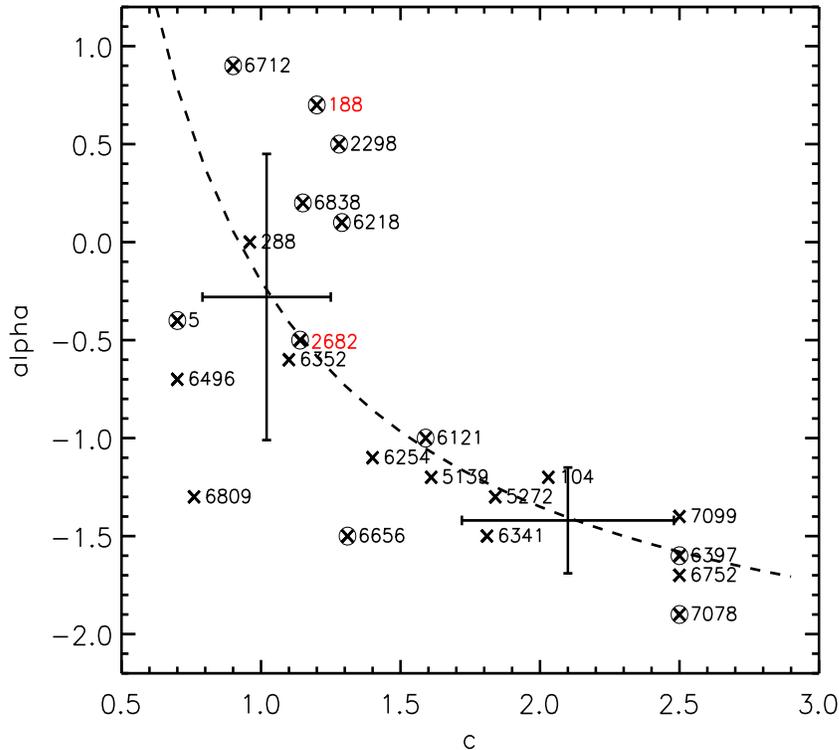}
\caption{Observed trend between MF index $\alpha$ and the central
concentration parameter $c$. Clusters are indicated by their NGC (or
Pal) index number. Objects for which a GMF index is available are
marked with a circled cross. For all others, the value of $\alpha$
is that of the MF measured near the half-light radius.}
\label{fig1}
\end{figure}

In Figure\,\ref{fig1} we show the run of the GMF index $\alpha$ as a 
function of $c$ for the clusters in our sample, as indicated by the labels.
Besides the 20 GCs mentioned above (see De Marchi, Paresce \& Pulone 2007
for details on the sample), we have also included in Figure\,\ref{fig1} 
the old open clusters NGC\,188 and M\,67 (NGC\,2682), whose values of $c$ 
and $\alpha$ (for stars in the range $0.6 - 1.2$\,\Msolar) were derived 
from the photometric studies of Stetson, McClure \& VandenBerg (2004) and 
Fan et al. (1996), respectively. 

It is immediately obvious that the data do not follow the expected 
correlation or trend between increasing central concentration and 
flattening GMF mentioned above, as there are no high-concentration 
clusters with a shallow GMF. The median value of $c$ ($1.4$) splits the 
cluster population roughly in two groups, one with lower and one with 
higher concentration. The mean GMF index of the first group is $\alpha=-0.3 
\pm 0.7$, while the second has a much tighter distribution with 
$\alpha=-1.4 \pm 0.3$ (see large thick crosses in Figure\,\ref{fig1}). The 
relationship $\alpha+2.5=2.3/c$, shown as a dashed line in Figure\,\ref{fig1},
is a simple yet satisfactory eye-ball fit to the distribution.

Figure\,\ref{fig1} suggests that a relatively low concentration is a 
necessary condition (although probably not also a sufficient one) for a 
depleted GMF. It appears, therefore, that mass loss, even severe, via
evaporation and tidal truncation has not triggered core-collapse for
low-concentration clusters (hence the title of this contribution). This
finding is unexpected and counter-intuitive. Although no satisfactory 
explanation presently exists for the observed behaviour, we briefly 
address here below some hypotheses. Some of the ideas put forth here 
are still preliminary, as they emerged from discussions during this 
Symposium.

IMF variations could explain the existence of clusters with a very
depleted GMF and a loose core. Some clusters with shallow IMF have 
undergone severe stellar mass-loss and have therefore expanded 
considerably. This has led to a lower $c$ and a shallower GMF because 
puffed-up systems of this type were more prone to tidal truncation. 
Most of these clusters have already disrupted but some survive for a 
long time in a state of low $c$ and large $\alpha$. This hypothesis, 
however, does not explain the absence of clusters with a dense core 
and a shallow GMF, as the most massive clusters with an originally 
shallow IMF should have long collapsed and still be visible in the 
upper-right portion of Figure\,\ref{fig1}.

Alternatively, it is possible that the clusters with a depleted GMF 
and a loose core have in fact undergone core-collapse and have already 
recovered a normal radial density and surface brightness profile. Core 
re-expansion thanks to the energy released by hardening binaries has long 
been predicted (Hut 1985). However, the timescale for re-expansion, at 
least according to the predictions of Murphy, Cohn \& Hut (1990), seems 
too long to account for the observed distribution. It is more likely
that burning of primordial binaries may have halted the collapse
altogether. Calculations by Trenti (2007; 2008) suggest that a binary 
fraction $\gtrsim 10$\,\% in the core of loose clusters could be 
sufficient to avoid their collapse, without preventing further mass loss 
via evaporation or tidal stripping. This explanation seems particularly 
appealing in light of the discovery that many loose clusters appear to 
have a significant ($\gtrsim 6$\,\%) binary fraction in their cores 
(Sollima et al. 2007). However, the problem remains that some of the 
most depleted objects in the upper-left quadrant of Figure\,\ref{fig1} 
have too narrow a main sequence in the colour--magnitude diagram to 
account for a binary fraction in excess of a few percent (Pulone \& 
De Marchi, in preparation; see also Davis \& Richer 2008).

Another explanation for the depleted clusters with a loose core is that
proposed by Kroupa (2008) and Baumgardt (2008), in which the low 
concentration is the result of rapid gas expulsion from the cores of 
primordially segregated clusters in the early phases of their lives. Such 
a violent process would deplete the low-mass end of the GMF and could 
leave the GCs in an almost collisionless state, in which further dynamical 
evolution via two-body relaxation is prevented. The problem with this 
scenario, however, is how to explain why even the most depleted clusters 
such as NGC\,2298 or NGC\,6218 are in a condition of energy equipartition 
(De Marchi \& Pulone 2007; De Marchi et al. 2006), unless the observed 
mass stratification is the residue of primordial segregation.   

The apparently simple dependence of $\alpha$ from $c$ in Figure\,\ref{fig1}
might also suggest that the observed distribution in practice represents
an evolutionary sequence. In this scenario, the value of $c$ at the time
of cluster formation determines its evolution along two opposite directions 
of increasing and decreasing concentration. Clusters born with sufficiently 
high concentration ($c \gtrsim 1.5$) evolve towards core-collapse. Mass 
loss can be important via stellar evolution in the first $\sim 1$\,Gyr, and 
to a lesser extent via evaporation or tidal stripping throughout the life 
of the cluster, but the GMF at any time does not depart significantly from
the IMF. Clusters with $c \lesssim 1.5$ at birth also evolve towards 
core-collapse, but mass loss via stellar evolution and, most importantly,
via relaxation and tidal stripping proceeds faster, particularly if their
orbit has a short perigalactic distance or frequent disc crossings.
Therefore, as the tidal boundary shrinks and the cluster loses
preferentially low-mass stars, the GMF progressively flattens. This speeds 
up energy equipartition, but $c$ still decreases, since the tidal radius 
shrinks more quickly than the luminous core radius (although the central 
density, particularly that of heavy remnants, is increasing). These 
clusters could eventually undergo core-collapse, but this might only affect 
a few stars in the core, thereby making it observationally hard to detect. 
The signature of core-collapse might only be present and should therefore 
be searched in the radial distribution of heavy remnants (Mark Gieles, 
private communication).

In summary, while no conclusive explanation still exists for the unexpected 
observed trend between central concentration and shape of the GMF, 
Figure\,\ref{fig1} should serve as a warning that the surface brightness 
profile and the central concentration parameter of GCs are not as reliable 
indicators of their dynamical state as we had so far assumed. In fact, if a 
central cusp in the surface brightness profile were the signature of a 
cluster's post core-collapse phase, it would be hard to explain why only 
about 20\,\% of the Galactic GCs show a cusp when the vast majority of them 
are an order of magnitude older than their half-mass relaxation time (Ivan 
King, private communication). Our current estimate of the fraction of post 
core-collapse clusters may therefore need a complete revision as a large 
number of them may be lurking in the Milky Way. A more reliable assessment 
of a cluster's dynamical state requires the study of the complete radial 
variation of its stellar MF and of the properties of its stellar population, 
particularly in the core.

\begin{acknowledgments}

We are grateful to Enrico Vesperini, Pavel Kroupa, Holger Baumgardt, 
Michele Trenti, Simon Portegies Zwart, Sidney van den Bergh, Oleg Gnedin, 
Mark Gieles and Evghenii Gaburov for helpful discussions and suggestions.

\end{acknowledgments}


\begin{thebibliography}{}

\bibitem[Andreuzzi et al.(2001)]{and01} Andreuzzi, G., De Marchi, G., 
  Ferraro, F., Paresce, F., Pulone, L. 2001, A\&A, 372, 851
\bibitem[Baumgardt(2008)]{bau08}Baumgardt, H. 2008, these proceedings
\bibitem[Da Costa(1982)]{dac82} Da Costa, 1982, AJ, 87, 990
\bibitem[Davis \& Richer(2008)]{dav08} Davis, S., Richer, H. 2008, these
  proceedings
\bibitem[De Marchi et al.(1999)]{dem99} De Marchi, G., Leibundgut, B.,
  Paresce, F., Pulone, L. 1999, A\&A 343, 9L
\bibitem[De Marchi \& Paresce(1995)]{dem95} De Marchi, G., Paresce, F.
  1995, A\&A, 304, 202
\bibitem[De Marchi, Paresce \& Portegies Zwart(2005)]{dem05} De Marchi, G.,
  Paresce, F.,  Portegies Zwart, S. 2005, in ASSL 327, The initial mass
  function 50 years later, Eds. E. Corbelli,  F. Palla, H. Zinnecker
  (Dordrecht: Springer), 77
\bibitem[De Marchi, Paresce \& Pulone(2000)]{dem00} De Marchi, G.,
  Paresce, F., Pulone, L. 2000, ApJ, 530, 342
\bibitem[De Marchi, Paresce \& Pulone(2007)]{dem07b} De Marchi, G.,
  Paresce, F., Pulone, L. 2007, ApJ, 656, L65
\bibitem[De Marchi \& Pulone(2007)]{dem07a} De Marchi, G.,
  Pulone, L. 2007, A\&A, 467, 107
\bibitem[De Marchi, Pulone \& Paresce(2006)]{dem06} De Marchi, G.,
  Pulone, L., Paresce, F. 2006, A\&A, 449, 161
\bibitem[Djorgovski \& King(1986)]{djo86} Djorgovski, S., King, R. 1986,
  ApJ, 305, L61
\bibitem[Djorgovski \& Meylan(1993)]{djo93} Djorgovski, S., Meylan, G.
  1993, in ASP Conf. Ser. 50, Structure and Dynamics of Globular
  Clusters, S. Djorgovski, G. Meylan (San Francisco: ASP), 325
\bibitem[Elson, Hut \& Ingaki(1987)]{els87} Elson, R., Hut, P., Ingaki,
  S. 1987, ARAA, 25, 565
\bibitem[Fan et al.(1996)]{fan96} Fan, X., et al. 1996, AJ, 112, 628
\bibitem[Gnedin \& Ostriker(1997)]{gne97} Gnedin, O., Ostriker, J. 1997,
  ApJ, 474, 223
\bibitem[Harris(1996)]{har96} Harris, W. 1996, AJ 112, 1487
\bibitem[Hut(1985)]{hut85} Hut, P. 1985, in IAU Symp. 113, Dynamics of star
  clusters, (Dordrecht: Reidel), 231
\bibitem[Koch et al.(2004)]{koc04} Koch, A., Grebel. E., Odenkirchen, M.,
  Martinez--Delgado, D., Caldwell, J. 2004, AJ, 128, 2274
\bibitem[Kroupa(2008)]{kro08} Kroupa, P. 2008, these proceedings
\bibitem[Murphy, Cohn \& Hut(1990)]{mur90} Murphy, B., Cohn, H., Hut, P.
  1990, MNRAS, 245, 335
\bibitem[Paresce \& De Marchi(2000)]{pdm00} Paresce, F., De Marchi, G.
  2000, ApJ, 534, 870
\bibitem[Paresce, De Marchi \& Jedrzejewski(1995)]{par95} Paresce, F.,
  De Marchi, G., Jedrzejewski, R. 1995, 442, L57
\bibitem[Richer et al.(1991)]{ric91} Richer, H., Fahlman, G., Buonanno, 
  R., Fusi Pecci, F., Searle, L., Thompson, I. 1991 381, 147
\bibitem[Sirianni et al.(2002)]{sir02} Sirianni, M., Nota, A., De Marchi, 
  G., Leitherer, C., Clampin, M. 2002, ApJ, 579, 275
\bibitem[Sollima et al.(2007)]{sol07} Sollima, A., Beccari, G., Ferraro, 
  F., Fusi Pecci, F., Sarajedini, A. 2007, MNRAS, 380, 781
\bibitem[Spitzer(1987)]{spi87} Spitzer, L. 1987, Dynamical evolution of
  globular clusters, (Princeton: Princeton Univ. Press) 
\bibitem[Stetson, McClure \& VandenBerg(2004)]{ste04} Stetson, P., McClure,
  R., VandenBerg, D. 2004, PASP, 116, 1012
\bibitem[Trenti(2007)]{tre07} Trenti, M. 2007, American Astronomical 
  Society, DDA meeting \#38, \#2.01
\bibitem[Trenti(2008)]{tre08} Trenti, M. 2008, these proceedings
\bibitem[Trager, Djorgovski \& King(1995)]{tra95} Trager, S., Djorgovski,
  S., King, I. 1995, AJ, 109, 218
\bibitem[Vesperini \& Heggie(1997)]{ves97} Vesperini, E., Heggie, D. 1997,
  MNRAS, 289, 898

\end{thebibliography}
\end{document}